\documentclass[conference]{IEEEtran}
\IEEEoverridecommandlockouts
\usepackage{cite}
\usepackage{amsmath,amssymb,amsfonts}
\usepackage{algorithmic}
\usepackage{graphicx}
\usepackage{textcomp}
\usepackage{xcolor}
\usepackage{comment}
\usepackage{url}

\usepackage{caption}
\usepackage{subcaption}
\usepackage{multirow}
\usepackage{multicol}
\usepackage{array}

\def\BibTeX{{\rm B\kern-.05em{\sc i\kern-.025em b}\kern-.08em
    T\kern-.1667em\lower.7ex\hbox{E}\kern-.125emX}}

\makeatletter 
\newcommand{\linebreakand}{%
  \end{@IEEEauthorhalign}
  \hfill\mbox{}\par
  \mbox{}\hfill\begin{@IEEEauthorhalign}
}

\makeatother 

\begin{document}

\title{Exploring Feature Importance and Explainability Towards Enhanced ML-Based DoS Detection in AI Systems \\

}

\author{\IEEEauthorblockN{Paul Badu Yakubu}
\IEEEauthorblockA{\textit{Computer \& Information Science} \\
\textit{Fordham University}\\
NY, NY USA \\
pbaduyakubu@fordham.edu}
\and
\IEEEauthorblockN{Evans Owusu}
\IEEEauthorblockA{\textit{Computer \& Information Science} \\
\textit{Fordham University}\\
NY, NY USA \\
eowusu3@fordham.edu}
\and
\IEEEauthorblockN{Lesther Santana}
\IEEEauthorblockA{\textit{Computer \& Information Sciences} \\
\textit{Fordham University}\\
NY, USA \\
lsantanacarmona@fordham.edu}
\and
\IEEEauthorblockN{Mohamed Rahouti}
\IEEEauthorblockA{\textit{Computer \& Information Science} \\
\textit{Fordham University}\\
NY, NY USA \\
mrahouti@fordham.edu}
\and

\IEEEauthorblockN{Abdellah Chehri}
\IEEEauthorblockA{\textit{Mathematics \& Computer Science} \\
\textit{Royal Military College of Canada}\\
Kingston, Ontario, Canada \\
chehri@rmc.ca}
\and


\and
\IEEEauthorblockN{Kaiqi Xiong}
\IEEEauthorblockA{\textit{Cyber Florida} \\
\textit{University of South Florida}\\
Tampa, FL USA \\
xiongk@usf.edu}
}

\maketitle

\begin{abstract}
Denial of Service (DoS) attacks pose a significant threat in the realm of AI systems security, causing substantial financial losses and downtime. However, AI systems' high computational demands, dynamic behavior, and data variability make monitoring and detecting DoS attacks challenging. Nowadays, statistical and machine learning (ML)-based DoS classification and detection approaches utilize a broad range of feature selection mechanisms to select a feature subset from networking traffic datasets. Feature selection is critical in enhancing the overall model performance and attack detection accuracy while reducing the training time. In this paper, we investigate the importance of feature selection in improving ML-based detection of DoS attacks. Specifically, we explore feature contribution to the overall components in DoS traffic datasets by utilizing statistical analysis and feature engineering approaches. Our experimental findings demonstrate the usefulness of the thorough statistical analysis of DoS traffic and feature engineering in understanding the behavior of the attack and identifying the best feature selection for ML-based DoS classification and detection.

\end{abstract}


\section{Introduction} Internet Service Providers (ISPs) around the globe continue to experience a dramatic growth in network attacks. Notably, the Denial of Service (DoS) is considered one of the most harmful attacks on networked systems. Such attacks cost an average of \$22,000 for every minute of downtime they cause, resulting in an average loss of \$120,000 per attack for small to medium-sized businesses \cite{DoS-harm}. Therefore, understanding DoS traffic behaviors and containing DoS attacks become necessary and substantial.

Under DoS attacks, the network can become overwhelmed by the burst of traffic, rendering it unavailable to process legitimate service requests from clients or other network entities. DoS type of network anomalies is classified into exploitation and reflection attacks. The escalation of these anomalies on the Internet has remarkably shifted the attention of researchers/service providers to explore network threats deeply \cite{david2021discriminating}. Various studies have been conducted to investigate network anomalies' potency and propose novel detection and mitigation mechanisms in the past \cite{rahouti2021synguard, zheng2018realtime}. However, a noticeable problem remains challenging in this direction. It is related to the stealthiness of application layer DoS attacks, as this type of anomaly does not typically manifest at the network level, allowing these anomalies to evade the conventional network layer detection solutions. It is crucial to have an in-depth perception of network traffic behaviors and promptly detect and control network DoS attacks. Thus, this work tries to analyze the traffic behaviors of various DoS anomalies and distinguish them from normal network traffic for further lightweight detection. 

Specifically, we aim to leverage component analysis and ML techniques for efficient traffic analysis and detection of DoS attacks, the dominant security threat to the Internet. Most existing efforts rely on threshold-based heuristics and/or simple statistics, which cannot meet the real-time requirement of Internet systems, especially facing the ever-growing number and scales of different types of DoS attacks. We propose to fill this gap by properly deploying feature selection with efficient exploratory analysis of DoS behavior for resource-efficient characterization of Internet traffic flows. That is, we present an efficient methodology for network traffic flow characterization and classification models based on a broadly-used statistical method and new traffic data screening and selection. In addition to the commonly used flow-level features, we consider more informative metrics that capture traffic flow's volume and velocity features to improve the analysis's accuracy and precision.

The rest of this paper is organized as follows. Section \ref{sec:background} discusses the research problem background and Section \ref{sec:related} summarizes the literature review. Next, the methodology used in this work is presented in Section \ref{sec:methodology}. Section \ref{sec:evaluation} presents the evaluation setups and key findings. Last, Section \ref{sec:conclusion} concludes this work. \label{sec:introduction}

\section{Research Problem and Goals} DoS attacks remain a significant threat to computer networks, causing disruption and loss of service. ML techniques have been widely used to detect and prevent these attacks. However, the large volume and high-dimensional nature of network traffic data present a challenge for accurately and efficiently detecting DoS attacks using these ML algorithms.

In this work, we aim to investigate the effectiveness of using Principal Component Analysis (PCA) procedures to filter out irrelevant features from the LYCOS-IDS2017 dataset \cite{lycos} before training ML models for detecting DoS attacks. We hypothesize that using PCA to filter out irrelevant features from the dataset will improve the accuracy of the ML models while reducing the computational cost. Additionally, we expect that by comparing the performance of different ML models, including decision trees (DTs), random forests (RFs), and support vector machines (SVMs), we can identify the most effective model for DoS attack detection.

Further, finding highly relevant/important features to exploring DoS patterns in the network dataset is vital. Therefore, we must examine the highly pertinent features for each flow class label and discard irrelevant (low-importance) features. This can assist in training the classification model in a lightweight setup (lightweight classifier) with a small set of features while alleviating the overfitting problem.

The overall research goal is to explore how feature importance and explainability can help enhance the learning model performance and DoS detection accuracy and how to optimally analyze the behavior of DoS events based on dataset component comparisons. For this goal, the dataset must first be preprocessed by removing irrelevant features and then conduct a comprehensive exploratory analysis of feature importance. We will next train multiple ML models, including DTs, RFs, and support vector machines, to detect DoS attacks using the preprocessed dataset. We will evaluate the performance of the trained models using various metrics such as accuracy, precision, recall, and F1 score.

We will further compare the performance of our methods with other state-of-the-art ones for detecting DoS attacks, including ML techniques that do rely on component analysis procedures for feature selection. By comparing the results, we will identify the most effective method for detecting DoS attacks using the LYCOS-IDS2017 dataset. The objectives of this work are (1) reducing the networking traffic dataset dimensionality, (2) alleviating the selection of correlated features (selecting only the important features that capture the majority of the variation in the data), and (3) enhancing the results' interpretability and generalizability (by reducing the risk of overfitting and improving the ability of ML models to generalize to new and unseen traffic patterns). The outcome of this work can provide a better understanding of the effectiveness of using component analysis procedures to filter out irrelevant features from the network traffic datasets for anomaly detection. 

\begin{table}[!ht]
\caption{Types of network traffic in the LYCOS-IDS dataset and their training and testing sets.}
\label{tab:lycos3}
\centering
\begin{small}
\begin{tabular}{llll}
09/25-10/09 & Coll \\
DoS\_hulk & 79,494 & 39,747 & 39,747 \\
Portscan & 79,465 & 39,732 & 39,732 \\
DDoS & 47,841 & 23,920 & 23,920 \\
DoS\_goldeneye & 3,382 & 1,691 & 1,691 \\
DoS\_slowloris & 2,837 & 1,418 & 1,418 \\
DoS\_slowhttptest & 2,433 & 1,216 & 1,216 \\
FTP\_patator & 2,001 & 1,000 & 1,000 \\
SSH\_patator & 1,479 & 739 & 739 \\
Webattack\_bruteforce & 680 & 340 & 340 \\
Bot & 367 & 183 & 183 \\
Webattack\_xss & 326 & 163 & 163 \\
Webattack\_sql\_injection & 6 & 3 & 3 \\
Heartbleed & 5 & 2 & 2 \\
\hline
Total & 440,632 &  220,312 &  220,312 \\
\end{tabular}
\end{small}
\end{table}

 \label{sec:background}

\section{Related Works} The escalation of DoS attacks on networks has significantly shifted the attention of researchers/service providers to investigate network-based DoS events intensively \cite{david2021discriminating}. Several studies have been carried out to explore the potency of DoS events and to introduce novel detection and mitigation approaches in the past years \cite{chin2018kernel, zheng2018realtime, rahouti2021synguard, liang2019empirical}. Further, DoS event traffic does not usually manifest at the network level, and they, therefore, can easily evade the traditional network layer detection methods.

On the one hand, feature engineering can implicitly impact the DoS events detection efficiency \cite{thakkar2021attack}. Thus, the effect of feature/component analysis procedures on Internet traffic is worth exploring as most of the statistical and ML solutions typically do not rely on exploratory analyses of feature importance, thus introducing distortion in the traffic, degrading the detection accuracy and capability to monitor the traces of DoS attacks \cite{jazi2017detecting}. On the other hand, understanding the statistical similarities in Internet flows can help differentiate DoS attacks from flush crowd events. Several existing studies tried to address such traffic characterization by utilizing discrimination techniques based on the flow statistical metric similarity obtained from the collected traffic \cite{yu2011discriminating}.

Further, feature selection is crucial in real network traffic, especially in high-dimensional network flows. A large number of Internet traffic features can lead to lengthy, resource and time-consuming training processes (especially under the presence of highly-correlated features), while the prediction accuracy does not consequently improve \cite{di2021supervised}. Several studies investigated feature selection approaches for DoS event detection. Among them, Saha et al. \cite{saha2022towards} evaluated the performance of feature selection approaches in DoS detection using deep learning and ML models. Sarhan et al. \cite{sarhan2022evaluating} examined essential feature sets to improve the generalisability of ML-based anomaly classification. Sanchez et al. \cite{sanchez2021feature} leveraged a feature selection approach for developing a deep learning-based DoS detection. Setitra et al. \cite{setitra2022feature} studied dimensionality reduction and feature modeling to enhance ML-based DoS identification in software-defined networks (SDN). Statistical characteristics of the correlation of the components were also investigated.

Several limitations remain in the previous works, such as a limited explainability of feature importance in identifying DoS patterns. Thus, this work proposes to fill this gap by efficiently employing statistical-based feature selection with efficient exploratory analysis of DoS behavior for resource-efficient characterization of Internet traffic flows, hence improving the ML-based detection of DoS attacks. This work further enriches the literature by demonstrating the impact of feature selection on ML solutions for network anomaly detection. \label{sec:related}

\section{Methodology} \label{sec:methodology} 

\subsection{Dataset}

We used the LYCOS-IDS2017 datasets \cite{rosay2021cic} created by LycoSTand flow extractor \cite{lycos}, which consists of five datasets corresponding to five days of the week. The dataset comprises 1,837,498 entries, each representing a network flow with 83 features. To split the dataset, we followed the same approach as the Lycos authors, which involves building training and test sets by randomly selecting 50\% of each attack for the training set, 25\% for the cross-validation set, and 25\% for the test set. Table \ref{tab:lycos3} provides the resulting attack distribution and Table \ref{tab:data-features} summarizes the flow statistics and features of the dataset. 

\begin{table}[h]
\centering
\caption{Flow features and statistics in the LYCOS-IDS dataset.} 
\label{tab:data-features}
\begin{tabular}{|p{2.5cm}|p{5.2cm}|} \hline
Feature & Info  \\ 
 \hline
Decision tuple & ID, src/dest IP, src/dst port, protocol \\ \hline
Time &  Timestamp, duration \\ \hline
Fwd pkts & Total, len (total, max, min, std, min) \\ \hline
Bwd pkts & Total, len (total, max, min, std, min)  \\ \hline
IAT & Mean, std, max, min \\ \hline
Fwd IAT & Total, mean, std, max, min \\ \hline
Bwd IAT & Total, mean, std, max, min \\ \hline
Fwd flags & Push, URG \\ \hline
Flags & Bwd (Push, URG), Count \\ \hline
Pkts len/size & Pkts (min, std, max, mean, var), size (avg) \\ \hline
Pkt loss & Down/up ratio \\ \hline
Flags count & FIN/SYN/RST/PSH/ACK/URG/CWE/ECE   \\ \hline
Fwd pkt header & Len, avg (Seg size, bytes/bulk, bulk rate) \\ \hline
Bwd pkt header & Len, avg (Seg size, bytes/bulk, bulk rate)     \\ \hline
Subflow & Fwd/bwd (pkts, Bytes) \\ \hline
Init win Bytes & Fwd, Bwd    \\ \hline
Active/idle & Mean, max, std, min  \\ \hline
Other labels & Inbound, Similar HTTP \\ \hline
\end{tabular}
\end{table}

The selection of the LYCOS-IDS2017 dataset over other datasets, such as CIC-IDS2017 \cite{sharafaldin2018toward}, is justified in \cite{rosay2022network}. One key advantage is the efficient handling of TCP session terminations during flow and packet processing. In UDP, the end of a flow is determined by a timeout, which is comparable to the way some TCP communications end. In typical TCP connections, flow duration is calculated between the synchronization (SYN) and finish (FIN) flags. In the LycoSTand flow extractor, UDP flows are closed with a configurable timeout of 120s, whereas TCP flows are closed as soon as a packet with reset (RST) or FIN flags are detected. The flowID is then added to the ``TCP terminated" list. When a new TCP packet arrives, it checks if the flowID is in the TCP terminated list. If it is, LycoSTand drops packets associated with that session termination until a new flow with the SYN flag is detected, which starts a new session. The TCP terminated list is then updated with each packet by removing packets older than the maximum duration, calculated using timestamps \cite{rosay2022network}. 

\subsection{Data Pre-processing and Exploratory Analysis}

PCA is used to reduce the dimensionality of a dataset by transforming the original features into a new set of uncorrelated features, aka principal components. Reducing the number of variables in our dataset may lead to a loss in accuracy. Still, the aim is to trade some accuracy for simplicity, which speeds up the analysis of data points for ML algorithms. The goal is to establish a smaller dataset that is easier to work with and still retains much of the original traffic information. 

Further, since all entries were initially null in the output flow file, we removed the three URG flag-related in addition to the flow characterization features (id, src and dest IPs and port, and protocol). Next, since we had ample feature space, we used PCA for dimensionality reduction. PCA has a tunable number of components to keep, determining the variance retained in the data. We utilized a pipeline of Z-score normalization and PCA with a DT using the training dataset to find the number of components to maintain. We measured how model performance varied by the number of components to keep using five-fold cross-validation. Building upon this, we computed the accuracy, precision, recall, F1, false positive rate (FPR), and training and inference for different targets of variance explained by the PCA components (50\%, 60\%, 70\%, 80\%, 90\%, 95\%, and 99\%). Note that FPR is the percentage of benign traffic classified as attacks.

\begin{table}[h]
\centering
\caption{Model performance by variance explained by different PCA components. VE = variance explained percentage; \#C = number of components; T \& I = training \& inference Time (s); A = accuracy; P = precision; R = recall; FPR = percentage of false positive rate.} 
\label{tab:pca1}
\begin{tabular}{|p{0.6cm}|p{0.5cm}|p{1cm}|p{0.6cm}|p{0.6cm}|p{0.6cm}|p{0.6cm}|p{0.6cm}|}
\hline
 VE & \#C & T \& I & A & P & R & F1 & FPR \\ 
 \hline
50\% & 4 & 4.19 & 99.49 & 99.41 & 99.49 & 99.44 & 0.58 \\ \hline
60\% & 5 & 5.31 & 99.49 & 99.58 & 99.49 & 99.5 & 0.26 \\ \hline
70\% & 8 & 6.58 & 99.57 & 99.65 & 99.57 & 99.56 & 0.23 \\ \hline
80\% & 12 & 8.74 & 99.55 & 99.62 & 99.55 & 99.54 & 0.31 \\ \hline
90\% & 19 & 13.66 & 99.73 & 99.72 & 99.73 & 99.71 & 0.16 \\ \hline
95\% & 25 & 18.48 & 99.74 & 99.73 & 99.74 & 99.72 & 0.16 \\ \hline
99\% & 36 & 24.72 & 99.72 & 99.7 & 99.72 & 99.7 & 0.21 \\ \hline
\end{tabular}
\end{table}

Table \ref{tab:pca1} presents the model performance by variance explained by different PCA components. As the target percentage of variance explained by the PCA components increases, the model's performance metrics generally improve, except for the inference time, which also increases. The model's accuracy, precision, recall, and F1 score all increase from 50\% to 90\% variance explained, with the highest values achieved at 90\% variance explained. FPR decreases as the target percentage of variance explained increases, indicating that less benign traffic is classified as an attack. However, when the target percentage of variance explained is set to 95\%, the model's performance metrics slightly decrease, compared to the results obtained with 90\% variance explained. This could indicate that the model's complexity increases when more principal components are included, leading to overfitting and decreased performance. Overall, the table demonstrates the trade-off between model performance and computational complexity in the context of dimensionality reduction with such a component analysis procedure.

To better understand how the original feature space influences the components, we take eigenvectors associated with the top three components, which amount to about 48\% of the variance. The values in the eigenvectors represent the relative importance of each original feature in creating the principal component. By examining the magnitude and sign of the values, we can understand how much that feature contributes to each component. Larger magnitudes indicate more substantial contributions. Based on the heatmap examination of the features that exert the most influence, component 1's most critical features are flow duration and two features group. One has features with descriptive statistics (min, max, mean, standard deviation) of time between packets in a flow. They are iat\_max, iat\_mean, iat\_std, fwd\_iat\_tot, fwd\_iat\_max, fwd\_iat\_min, fwd\_iat\_mean, fwd\_iat\_std, bwd\_iat\_tot, bwd\_iat\_max, bwd\_iat\_mean, bwd\_iat\_std.

The second group has features that measure flow idle time. Among them, idle\_max, idle\_min, idle\_mean. As for the second component, three groups of features measure flow packet count, bytes used for packet headers, packet size, protocol flags, sub-flow statistics, and non-empty TCP packets. For the third component, three groups incur the most significance, attributes with packet length descriptive statistics, protocol flags, and data transmitted during TCP initial connection setup. To gain insights into how the original feature space affects the components, we examine eigenvectors related to the top three components, which account for approximately 48\% of the variance. These eigenvectors represent the relative importance of each original feature in creating the principal component. We can determine how much each feature contributes to each component by analyzing the magnitude and sign of the values. Larger magnitudes indicate more significant contributions.

Component 1, which accounts for the largest portion of the variance, is primarily influenced by flow duration and two feature groups. The first group includes features that describe the time between packets in a flow, such as iat\_max, iat\_mean, and iat\_std. These features are particularly useful for detecting network attacks that involve abnormal packet timing behavior, such as DDoS attacks or port scanning. The second group in Component 1 includes features that measure flow idle time, such as idle\_max, idle\_min, and idle\_mean. These features can help detect network attacks that involve periods of inactivity, such as some types of DoS attacks.

Component 2 is heavily influenced by three feature groups that measure flow packet count, bytes used for packet headers, packet size, protocol flags, sub-flow statistics, and non-empty TCP packets. These features are particularly useful for detecting attacks that involve abnormal packet behaviors, such as network scanning and some types of DoS attacks. Finally, Component 3 is primarily influenced by three feature groups, including packet length descriptive statistics, protocol flags, and data transmitted during TCP initial connection setup. These features are beneficial for detecting network attacks involving abnormal packet sizes, such as network scanning and data exfiltration attacks.

\subsection{Dimensionality Reduction using DT}

Separately identifying important features relevant to DoS patterns in the dataset is crucial. Here, we aim to explore the highly pertinent features for each flow class label and discard irrelevant (low-importance) ones. As a result, this helps train the classification model with a small set of features (lightweight classifier) while alleviating the overfitting problem.

\begin{figure}[h]
	\centering
	\includegraphics[height=7cm, width=8.8cm]{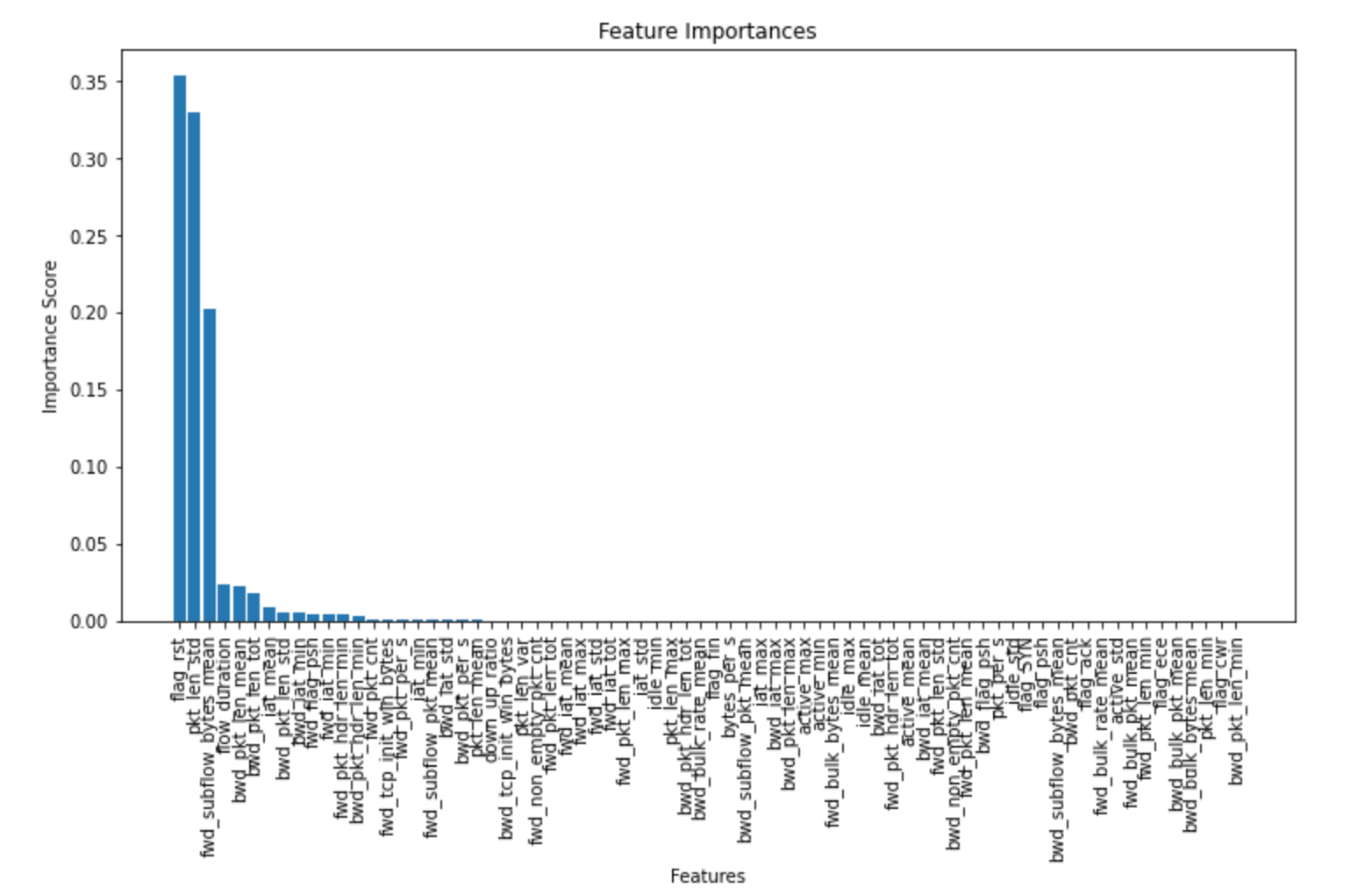}
	\caption{Top important features by DT.}
	\label{fig:dt_feature_importance}
\end{figure}

The analysis of feature importance using a DT, as shown in Fig. \ref{fig:dt_feature_importance}, identifies the most crucial features in enabling the model to differentiate between benign traffic and DoS attacks accurately. Also, fitting the DT on top of these features produced an accuracy of 99.50 \%, precision of 99.50\%, recall of 99.50\%, F1 score of 99.50\%, and FPR of 4.3\%. The identified features are sorted in order of importance as follows. 1) flag\_rst: indicates the number of TCP RST packets in a flow used to terminate TCP connections; 2) pkt\_len\_std: indicates the standard deviation of packet lengths in the flow; 3) fwd\_subflow\_bytes\_mean: represents the mean number of bytes per subflow (a subset of packets in a flow sharing common characteristics, such as IP addresses and ports) in the forward direction; 4) flow\_duration: represents the flow duration, which can be used to identify short-lived flows-based attacks; 5) bwd\_pkt\_len\_mean: indicates the mean size of packets in the backward direction of a flow.; and 6) bwd\_pkt\_len\_tot: represents the total size of packets in the backward direction of a flow.

\section{Evaluation} \begin{figure*}
    \centering
    \begin{subfigure}[b]{0.48\textwidth}
        \includegraphics[height=5.4cm, width=8.4cm]{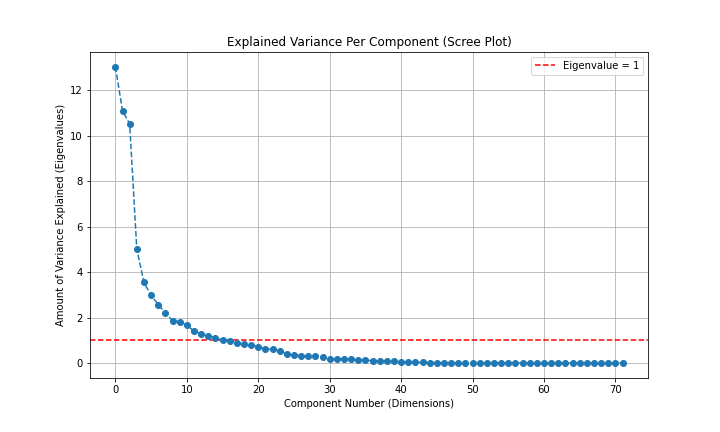}
        \caption{Explained variance per component using Scree Plot.}
        \label{fig:exp-var-comp}
    \end{subfigure}
    \begin{subfigure}[b]{0.48\textwidth}
        \includegraphics[height=5.4cm, width=8.4cm]{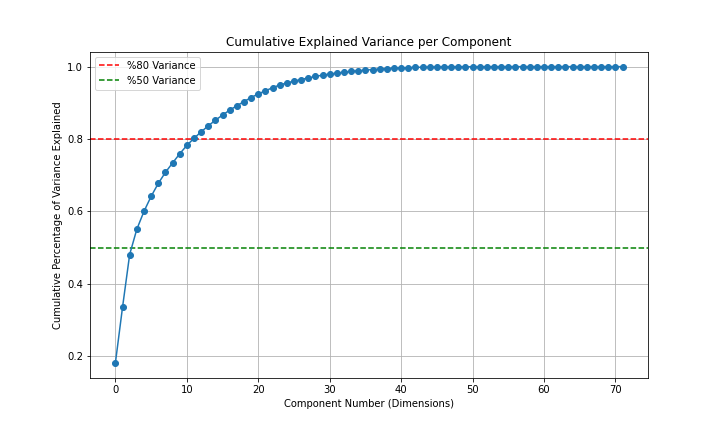}
        \caption{Cumulative explained variance per component.}
        \label{fig:cum-variance}
    \end{subfigure}
    \caption{Explained and cumulative explained per-component variance.}
    \label{fig...}
\end{figure*}

Figures \ref{fig:exp-var-comp} and \ref{fig:cum-variance} depict two ways to estimate how many components to keep. On the one hand, Fig. \ref{fig:exp-var-comp} plots the eigenvalues of each principal component to find the point where the slope decreases (or elbow) to avoid choosing the components that explain less significant additional variance. On the other hand, Fig. \ref{fig:cum-variance} gives the cumulative ratio of variance captured by each principal component sorted from highest to lowest to find how many to choose to retain a certain variance amount.

\begin{table}[h]
\centering
\caption{Model performance findings with PCA analyses. TT = training time (s); IT = inference time (s).} 
\label{tab:with-pca}
\begin{tabular}{|p{1.1cm}|p{0.5cm}|p{0.6cm}|p{0.6cm}|p{0.6cm}|p{0.6cm}|p{0.6cm}|}
\hline
 Algorithm & P & R & F1 & FPR & TT & IT \\ 
 \hline
LDA & 90.26 & 90.13 & 90.04 & 2.38 & 3.16 & 0.48 \\ \hline
QDA & 98.53 & 97.46 & 97.84 & 0.64 & 2.97 & 0.78 \\ \hline
SVM & 95.75 & 96.93 & 95.94 & 3.84 & 135.06 & 0.25 \\ \hline
k-NN & 99.79 & 99.8 & 99.79 & 0.06 & - & 6.06 \\ \hline
DT & 99.7 & 99.66 & 99.64 & 0.22 & 5.79 & 0.22 \\ \hline
RF & 99.8 & 99.81 & 99.8 & 0.1 & 29.06 & 1.44 \\ \hline
\end{tabular}
\end{table}

\begin{table}[h]
\centering
\caption{Model performance findings without PCA.} 
\label{tab:without-pca}
\begin{tabular}{|p{1.1cm}|p{0.5cm}|p{0.6cm}|p{0.6cm}|p{0.4cm}|p{0.6cm}|p{0.9cm}|}
\hline
 Algorithm & P & R & F1 & FPR & TT & IT \\ 
 \hline
LDA   & 97.98 & 95.96 & 96.82 & 0.44 & 5.51 & 0.31 \\ \hline
QDA   & 99.67 & 99.58 & 99.52 & 0.4 & 2.3 & 2.22 \\ \hline
SVM   & 99.28 & 99.36 & 99.23 & 0.81 & 283.1 & 0.25 \\ \hline
k-NN  & 99.82 & 99.82 & 99.82 & 0.04 & - & 3096.54 \\ \hline
DT    & 99.8 & 99.81 & 99.79 & 0.09 & 2.87 & 0.52 \\ \hline
RF    & 99.85 & 99.86 & 99.85 & 0.03 & 20.37 & 1.45 \\ \hline
\end{tabular}
\end{table}

\begin{table}[h]
\centering
\caption{Performance metrics for different attacks.} 
\label{tab:individual_attack_detection}
\begin{tabular}{|p{2.7cm}|p{1cm}|p{0.9cm}|p{0.9cm}|p{0.9cm}|}
\hline
 Attack & Precision & Recall & F1\-score & Support \\ 
 \hline
Benign & 1 & 1 & 1 & 110158 \\ \hline
DDoS & 1 & 1 & 1 & 23920 \\ \hline
DoS\_hulk & 1 & 1 & 1 & 39747 \\ \hline
Portscan & 1 & 1 & 1 & 39732 \\ \hline
DoS\_goldeneye & 0.95 & 0.98 & 0.96 & 1691 \\ \hline
DoS\_slowhttptest & 0.97 & 0.98 & 0.97 & 1216 \\ \hline
DoS\_slowloris & 0.98 & 0.98 & 0.98 & 1418 \\ \hline
SSH\_patator & 1 & 0.99 & 1 & 739 \\ \hline
FTP\_patator & 1 & 1 & 1 & 1000 \\ \hline
Webattack\_bruteforce & 0.7 & 0.81 & 0.75 & 340 \\ \hline
Bot & 0.97 & 0.98 & 0.98 & 183 \\ \hline
Webattack\_xss & 0.45 & 0.31 & 0.36 & 163 \\ \hline
Webattack\_sql\_injection & 0 & 0 & 0 & 3 \\ \hline
Heartbleed & 0.4 & 1 & 0.57 & 2 \\ \hline
\end{tabular}
\end{table}


We trained different ML algorithms using the top 12 PCA components. They explain around 80\% variance of the original dataset. Training time refers to the time it takes to fit the model to the training dataset, whereas inference time refers to the time it takes to use the trained model to make predictions on data points. As this is an imbalanced multilabel classification, we calculate metrics for each label and determine the average by support weight. The weighted recall is equivalent to accuracy. Based on the obtained results shown in Table \ref{tab:with-pca}, k-NN achieved the highest precision, recall, and F1 score, with an accuracy of 99.80\% and an FPR of only 0.06\%. DT and RF also achieved high accuracy scores of 99.66\% and 99.81\%, respectively, but with higher FPRs compared to k-NN.

Further, the linear discriminant analysis (LDA) achieved an accuracy of 90.13\%, the lowest among all the algorithms. The quadratic discriminant analysis (QDA) and SVM achieved high accuracy scores of 97.46\% and 96.93\%, respectively, with SVM incurring the highest FPR of 3.84\%. Regarding computational efficiency, k-NN had the longest inference time (6.06 seconds), while SVM had the longest training time (135.68 seconds). DT has relatively larger training times compared to LDA and QDA. DT, LDA, QDA, and SVM had significantly short inference times, while RF had relatively longer inference times. Overall, the results suggest that k-NN performs the best in terms of accuracy and FPR, while LDA has the highest FPR and lowest accuracy.

The results presented in Table \ref{tab:with-pca} using the test dataset indicate that k-NN achieved the highest accuracy, precision, recall, and F1 score, with an accuracy of 99.80\% and an FPR of only 0.06\%. DT and RF algorithms also showed high accuracy scores of 99.66\% and 99.81\%, respectively, although with slightly higher FPRs than k-NN. In contrast, LDA showed the lowest accuracy score of 90.13\% among all algorithms. QDA and SVM algorithms achieved high accuracy scores of 97.46\% and 96.93\%, respectively, but SVM showed the highest FPR of 3.84\%. Regarding computational efficiency, k-NN had the longest inference time of 6.06 seconds, while SVM had the longest training time of 135.68 seconds. DT had relatively larger training times compared to LDA and QDA. DT, LDA, QDA, and SVM had significantly short inference times, while RF had relatively longer ones. In summary, k-NN showed the best accuracy and FPR, while LDA exhibited the lowest accuracy score and the highest FPR.

Table \ref{tab:without-pca} shows the models' performance without the PCA dimensionality reduction pipeline. Comparing it with Table \ref{tab:with-pca} and focusing on the FPR, we see better results when PCA is not used. For instance, the FPR of LDA decreased from 2.384\% to 0.44\%, QDA from 0.64\% to 0.40\%, k-NN from 0.06\% to 0.04\%, DT from 0.22\% to 0.09\%, and RF from 0.10\% to 0.03\% and SVM from 3.84\% to 0.81\%. Regarding the other metrics (precision, recall, and F1 score), all models showed performance improvements. For the inference time, the k-NN significantly increased to 3,064 from 6.06 seconds. As a lazy learning algorithm, k-NN has a slower inference process because the dataset has much higher features.

Table \ref{tab:individual_attack_detection} shows the performance metrics for different attacks using K-NN as the detection model. The table reports the precision, recall, F1 score, and support for each type of attack. The support column shows the number of instances for each attack in the dataset. K-NN delivers perfect detection performance for the four most common attacks, with precision, recall, and an F1-score of 1.0, meaning that K-NN can accurately detect these attacks without false positives or negatives. However, for the less common attacks, performance varies. For example, for the DoS\_goldeneye attack, K-NN has a precision of 95\%, which means that 5\% of the instances detected as DoS\_goldeneye were false positives. The recall is 98\%, indicating that the model correctly detected 98\% of the DoS\_goldeneye attacks in the dataset. 

Similarly, for the webattack\_bruteforce attack, K-NN has a lower precision of 70\%, indicating that 30\% of the instances detected as webattack\_bruteforce were false positives. The recall is 81\%, indicating that K-NN correctly detected 81\% of the webattack\_bruteforce attacks in the dataset. Finally, for the attacks Webattack\_xss, Webattack\_sql\_injection, and Heartbleed, K-NN's performance is poor, with precision and recall of 0 or close to 0. In conclusion, K-NN shows excellent performance for the most common attacks but struggles with less frequent attacks. This highlights the importance of using multiple detection models or developing specialized models for specific attacks to improve overall network security. Ultimately, the k-NN classification results in \ref{tab:with-pca} imply that using PCA for feature selection in network intrusion detection models can maintain performance with faster inference times, allowing the model to identify and respond to potential threats quickly.

 \label{sec:evaluation}

\section{Conclusion}  

This work presented exploratory strategies using component analysis processes to investigate the importance of traffic features. Such strategies further help identify the contribution of features to the overall ML-based models for DoS detection. This work can enrich the literature by demonstrating the impact of feature selection on ML solutions for network anomaly detection. The evaluation results showed the importance of exploring feature importance and explainability toward accuracy-improved ML-enabled DoS detection.
 \label{sec:conclusion}


\bibliographystyle{IEEEtran}

\begin{thebibliography}{10}
\providecommand{\url}[1]{#1}
\csname url@samestyle\endcsname
\providecommand{\newblock}{\relax}
\providecommand{\bibinfo}[2]{#2}
\providecommand{\BIBentrySTDinterwordspacing}{\spaceskip=0pt\relax}
\providecommand{\BIBentryALTinterwordstretchfactor}{4}
\providecommand{\BIBentryALTinterwordspacing}{\spaceskip=\fontdimen2\font plus
\BIBentryALTinterwordstretchfactor\fontdimen3\font minus \fontdimen4\font\relax}
\providecommand{\BIBforeignlanguage}[2]{{%
\expandafter\ifx\csname l@#1\endcsname\relax
\typeout{** WARNING: IEEEtran.bst: No hyphenation pattern has been}%
\typeout{** loaded for the language `#1'. Using the pattern for}%
\typeout{** the default language instead.}%
\else
\language=\csname l@#1\endcsname
\fi
#2}}
\providecommand{\BIBdecl}{\relax}
\BIBdecl

\bibitem{DoS-harm}
``How much will a \mbox{DDoS} attack cost your business?'' \url{https://www.cloudbric.com/blog/2021/01/business-ddos-attacks-damages-and-cost/}, accessed: 2022-02-09.

\bibitem{david2021discriminating}
J.~David and C.~Thomas, ``Discriminating flash crowds from \mbox{DDoS} attacks using efficient thresholding algorithm,'' \emph{JPDC}, vol. 152, pp. 79--87, 2021, Elsevier.

\bibitem{rahouti2021synguard}
M.~Rahouti, K.~Xiong, N.~Ghani, and F.~Shaikh, ``\mbox{SYNGuard}: Dynamic threshold-based \mbox{SYN} flood attack detection and mitigation in software-defined networks,'' \emph{IET Networks}, vol.~10, no.~2, pp. 76--87, 2021.

\bibitem{zheng2018realtime}
J.~Zheng, Q.~Li, G.~Gu, J.~Cao, m.~Yau, and J.~Wu, ``Realtime \mbox{DDoS} defense using \mbox{COTS SDN} switches via adaptive correlation analysis,'' \emph{TIFS}, vol.~13, no.~7, pp. 1838--1853, 2018, IEEE.

\bibitem{lycos}
\BIBentryALTinterwordspacing
``Intrusion detection evaluation dataset (lycos-ids2017),'' accessed: April, 2023. [Online]. Available: \url{https://lycos-ids.univ-lemans.fr}
\BIBentrySTDinterwordspacing

\bibitem{chin2018kernel}
T.~Chin, K.~Xiong, and M.~Rahouti, ``Kernel-space intrusion detection using software-defined networking,'' \emph{EAI Endorsed Transactions on Security and Safety}, vol.~5, no.~15, p.~e2, 2018.

\bibitem{liang2019empirical}
X.~Liang and T.~Znati, ``An empirical study of intelligent approaches to \mbox{DDoS} detection in large scale networks,'' in \emph{ICNC}.\hskip 1em plus 0.5em minus 0.4em\relax IEEE, 2019, pp. 821--827.

\bibitem{thakkar2021attack}
A.~Thakkar and R.~Lohiya, ``Attack classification using feature selection techniques: a comparative study,'' \emph{JAIHC}, vol.~12, pp. 1249--1266, 2021, Springer.

\bibitem{jazi2017detecting}
H.~H. Jazi, H.~Gonzalez, N.~Stakhanova, and A.~A. Ghorbani, ``Detecting \mbox{HTTP}-based application layer \mbox{DoS} attacks on web servers in the presence of sampling,'' \emph{Computer Networks}, vol. 121, pp. 25--36, 2017.

\bibitem{yu2011discriminating}
S.~Yu, W.~Zhou, W.~Jia, S.~Guo, Y.~Xiang, and F.~Tang, ``Discriminating \mbox{DDoS} attacks from flash crowds using flow correlation coefficient,'' \emph{TPDS}, vol.~23, no.~6, pp. 1073--1080, 2011, IEEE.

\bibitem{di2021supervised}
M.~Di~Mauro, G.~Galatro, G.~Fortino, and A.~Liotta, ``Supervised feature selection techniques in network intrusion detection: A critical review,'' \emph{Engineering Applications of Artificial Intelligence}, vol. 101, p. 104216, 2021, Elsevier.

\bibitem{saha2022towards}
S.~Saha, A.~T. Priyoti, A.~Sharma, and A.~Haque, ``Towards an optimal feature selection method for \mbox{AI}-based \mbox{DDoS} detection system,'' in \emph{CCNC}.\hskip 1em plus 0.5em minus 0.4em\relax IEEE, 2022, pp. 425--428.

\bibitem{sarhan2022evaluating}
M.~Sarhan, S.~Layeghy, and M.~Portmann, ``Evaluating standard feature sets towards increased generalisability and explainability of \mbox{ML}-based network intrusion detection,'' \emph{Big Data Research}, vol.~30, p. 100359, 2022.

\bibitem{sanchez2021feature}
O.~R. Sanchez, M.~Repetto, A.~Carrega, R.~Bolla, and J.~F. Pajo, ``Feature selection evaluation towards a lightweight deep learning \mbox{DDoS} detector,'' in \emph{ICC}.\hskip 1em plus 0.5em minus 0.4em\relax IEEE, 2021, pp. 1--6.

\bibitem{setitra2022feature}
M.~A. Setitra, I.~Benkhaddra, Z.~E.~A. Bensalem, and M.~Fan, ``Feature modeling and dimensionality reduction to improve \mbox{ML}-based \mbox{DDoS} detection systems in sdn environment,'' in \emph{ICCWAMTIP}.\hskip 1em plus 0.5em minus 0.4em\relax IEEE, 2022, pp. 1--7.

\bibitem{rosay2021cic}
\BIBentryALTinterwordspacing
A.~Rosay, F.~Carlier, E.~Cheval, and P.~Leroux, ``From {CIC-IDS2017} to {LYCOS-IDS2017}: A corrected dataset for better performance,'' in \emph{IEEE/WIC/ACM WI-IAT}.\hskip 1em plus 0.5em minus 0.4em\relax ACM, 2021, p.~6. [Online]. Available: \url{https://doi.org/10.1145/3486622.3493973}
\BIBentrySTDinterwordspacing

\bibitem{sharafaldin2018toward}
I.~Sharafaldin, A.~H. Lashkari, and A.~A. Ghorbani, ``Toward generating a new intrusion detection dataset and intrusion traffic characterization,'' in \emph{ICISSP}, Portugal, January 2018.

\bibitem{rosay2022network}
A.~Rosay, E.~Cheval, F.~Carlier, and P.~Leroux, ``Network intrusion detection: A comprehensive analysis of cic-ids2017,'' in \emph{ICISSP}.\hskip 1em plus 0.5em minus 0.4em\relax SCITEPRESS-Science and Technology Publications, 2022, pp. 25--36.

\end{thebibliography}

\end{document}